\documentclass[letter,11pt]{article}

\usepackage[parfill]{parskip}
\usepackage{graphicx,amsmath,amsfonts,algorithm}
\floatname{algorithm}{Protocol}
\usepackage{graphics}
\usepackage{amssymb}
\usepackage{amsmath, amsthm}
\usepackage{complexity}
\usepackage{hyperref}
\usepackage{color}
\usepackage{bbm}
\usepackage[affil-it]{authblk}
\usepackage[margin=1in]{geometry}
\usepackage[normalem]{ulem}
\usepackage[all]{xy}\CompileMatrices

\def\qed{$\Box$}

\def\EQ#1{\begin{eqnarray}#1\end{eqnarray}}

\def\bra#1{\langle#1{|}}

\def\op#1{\hat{#1}}
\def\ket#1{| #1 \rangle}

\def\bra#1{\langle #1 |}

\def\op#1#2{|#1\rangle\langle#2|}

\definecolor{darkgreen}{RGB}{10,150,50}

\def\qed{$\Box$}

\newtheorem{prop}{Proposition}\def\PRO{\begin{prop}}\def\ORP{\end{prop}}
\newtheorem{coro}{Corollary}\def\COR{\begin{coro}}\def\ROC{\end{coro}}
\newtheorem{theo}{Theorem}\def\TH{\begin{theo}}\def\HT{\end{theo}}
\def\TH{\begin{theo}}\def\HT{\end{theo}}
\newtheorem{defi}[prop]{Definition}\def\DE{\begin{defi}}\def\ED{\end{defi}}
\newtheorem{lemme}[prop]{Lemma}\def\LE{\begin{lemme}}\def\EL{\end{lemme}}
\newcommand{\AR}[2][c]{$$\begin{array}[#1]{lllllllllllllll}#2\end{array}$$}

\def\EQ#1{\begin{eqnarray}#1\end{eqnarray}}

\def\op#1{\hat{#1}}
\def\ket#1{| #1 \rangle}

\def\bra#1{\langle #1 |}

\def\op#1#2{|#1\rangle\!	\langle#2|}

\def\dm#1{\op{#1}{#1}}
\def\ora#1{\overrightarrow{#1}}

\newcommand{\djj}{d\kern-0.4em\char"16\kern-0.1em}

\def\bb#1{ \boxed{#1} }

\usepackage[mathscr]{euscript}

\newcommand{\df}{\mathrel{\mathop:}=}

\begin{document}
\allowdisplaybreaks[3]
\frenchspacing

\title{Quantum-enhanced \\ Secure Delegated Classical Computing}

\author[1,2,3]{Vedran Dunjko}
\author[1]{Theodoros Kapourniotis}
\author[1]{Elham Kashefi}
\affil[1]{School of Informatics, University of Edinburgh, UK}
\affil[2]{Division of Molecular Biology, Ru\djj er Bo\v{s}kovi\'{c} Institute, Zagreb, Croatia}
\affil[3]{Now at: Austrian Academy of Sciences, Innsbruck, Austria}

\maketitle
\begin{abstract}

We present a quantumly-enhanced protocol to achieve unconditionally secure delegated classical computation where the client and the server have both {their }classical and quantum {computing} capacity {limited}. We prove the same task cannot be achieved using only classical protocols. This extends the recent work of Anders and Browne on the computational power of correlations to a security setting. Concretely, we present how a client with access to a non-universal classical gate such as a parity gate could achieve unconditionally secure delegated universal classical computation by exploiting minim{al} quantum gadgets. In particular, unlike the universal blind quantum computing protocols, the restriction of the task to classical computing removes the need for a full universal quantum machine on the {side of the} server and make{s} these new protocols readily implementable with the currently available quantum technology in the lab. 

\end{abstract}

\section{Introduction}

The concept of delegated quantum computing is the quantum extension of the classical task of computing with encrypted data without decrypting them first. The fully homographic encryption (FHE) scheme of \cite{Gentry09} has resolved this 30 years open questions in the classical setting with computational security. On the other hand many quantum protocols \cite{Childs05,AS06,UBQC,Dorit,MDK11,Science,TomoTopo,DKL12,monlyAlice,FM12,Morimae12,monlyAlice,SKM13,MPF13,DFPR13,CVK13,GMMR13,RUV13,FBS14} address this challenge for a futuristic quantum client-server setting achieving a wide range of security, and {other} properties.  Among all these protocols a family of protocols known as the Universal Blind Quantum Computing (UBQC) \cite{UBQC,MDK11,Science,TomoTopo,DKL12,monlyAlice} is the optimal one in terms of the client's requirements. The key properties of these protocols  are given below.

\begin{itemize}
\item The security is unconditional.
\item Client's classical operation{s are} efficient in the size of desired computation that is $O(poly (n))$ where $n$ is the size of the desired computation (both input and operations) plus classical memory.
\item Client's quantum operations {are} minimal, that is the generation of a restricted family of single random qubit{s} without any need for quantum memory.
\item Client and server have one-way quantum communication of the size of the desired computation, that is $O(poly (n))$, where $n$ is the size of the computation (both input and operations).
\item Client and server have {a} two-way classical communication of size $O(poly (n))$ where $n$ is the size of the computation (both input and operations).
\item Server has a universal quantum computer of size  $O(poly (n))$, i.e. he is able to manipulate coherently the creation of an entangled state of size $O(poly (n))$.  
\end{itemize}

The central challenge of the above protocols is the requirements of a server with a universal quantum computer. The main contribution of this paper is the design of a family of secure delegated protocols for only classical computing, where the server needs only to manipulate {a} few qubits. In what follows we present the strategy for generating several such protocols where the following concrete improvement will be obtained.

\begin{itemize}
\item Client{'s} classical operations are restricted to application of only XOR operators, {the generation of classical random bits,} and read out only memory of size $O(poly (n))$ where $n$ is again the size of the desired computation (both input and operations).
\item Server has a simple quantum device to manipulate coherently the creation of an entangled state of at most {a} constant {number of} qubits.  
\end{itemize}

Our family of protocols will provide a non-universal client the possibility of unconditionally secure delegation of any classical comput{ation} to a remote server that has access to basic quantum gadgets, currently available in many scientific and commercial labs. It is important to not{e} that such a functionality {cannot} be achieved with {just purely} classical devices as we prove later. Moreover, we prove that our protocols are, {in a sense}, optimal in the quantum setting as one {cannot} further simplify the {protocol requirements}, and {achieve the same task using} quantum states {which are} independent from the input (impossibility of off-line quantum communication {protocols}). Furthermore the requirements of the client's devices are also minimal which could lead to the design of miniature devices far smaller than any full scale classical computer. In comparison, in FHE protocols, {the} security is conditional on computational assumptions, {the} client needs to be universal, as does server, while the overhead remains large. However the goal of FHE protocols is different {to ours}: client's problem is not universality, but the complexity of the computation and communication with the server that needs to be independent of the complexity of the delegated computation. An interesting open question for future work is whether a hybrid combination of two schemes could lead to a more efficient (both in terms of performance and security) delegated computing scheme.

The structure of the rest of this paper is as follows, in Section 2 we describe the general concept common to our protocols. Our main methods are presented in Sections 3 and 4, including a family of entangled-based and also single qubit-based protocols. In Section 5 we present our main no-go results: of the unfeasibility of {achieving the same task with} a {purely} classical non-universal client, and the proof that our protocols cannot be modified to have {just} off-line communication. Section 6 discusses possible applications.

\section{General Idea}

The general idea behind all our proposed schemes for the secure computation of the universal NAND gate is based on the following fact presented for the first time in \cite{AB08}. Let $M^0$ to denote a Pauli-$X$ measurement and $M^1$ a Pauli-$Y$, then the three qubit measurement  $M^a \;\otimes\; M^b \;\otimes\; M^{a \oplus b}$ of the GHZ state (denoted in this paper as $\ket \Psi$) computes NAND$(a,b)$. We then extended this idea that instead of switching the measurements one can simply apply the pre-rotation operation based on $a$, $b$ and $a\oplus b$ to the GHZ state and then the Pauli-$X$ measurements of all three qubits achieve the same task. So the client effectively chooses the measurement basis by this pre-rotation while hiding his secret information as it is done in the universal blind quantum computing \cite{UBQC}. The next trick is that additional random $Z$ gates hide the outcome while achieving the same task. Our final generalisation is to notice that the operations need not to be performed on a GHZ state but could be performed sequentially on one single qubit $\ket +$ state as well. In other words if we denote the $\pi/2$ rotation along the $Z$ axis by $S$ then we prove that the local operators of the form
\AR{
S^{\dagger ^a} S^{\dagger^b} S^{\dagger^{a\oplus b}}
}
encode the input of the client in the resource state while permitting the server to perform the {other} operations {required to compute the} NAND gate. Since all the information of the client is encoded in the phase of the states, additional randomly chosen $Z$ gates achieve a full one-time pad of the client's information, which can easily be decoded by the client by a bit-flip. Analogous effect is achieved in the case of the single qubit resource. We can then design a family of encryption protocols in which $S^\dagger$ rotations, parametrized by input bits $a$ and $b$, and the XOR of the same input bits, along with a $Z$-phase rotation parameterised by a single or a multitude of encryption bits chosen by the client, prepare a resource state such that no information about the input bits is accessible to the server from the encrypted resource states, and such that a fixed measurement of this resource state results in a one-time padded bit equal to the NAND of the input bits.

While in this paper we have presented specific protocols based on various manipulations of the single qubit $\ket +$ and three qubits entangled GHZ state, one could easily adapt these protocols to cover various encodings necessary for the specific noise model or available resources within a particular implementation platform.  

\section{Entangled-based Protocols}

There are three types of protocols that we introduce here, to address various implementation scenarios. These families achieve the same goal and differ only in the required quantum gadgets of the client. In the first family  of the protocols, it is assumed that client {can} create or have secure access to some simple (few qubits) entangled states. On the other hand, in the second family it is assumed that the client is able to measure the flying qubit that it receive{s} through an untrusted channel to perform its desired universal comput{ation}. In the third setting, the client needs only to have the capacity to perform simple single qubit rotations. Importantly, in all three scenarios the classical computation of the client is restricted to XOR operations. 

\subsection{Preparing Client}

\begin{figure*}[h!] 
   \centering
   \includegraphics[width=300pt]{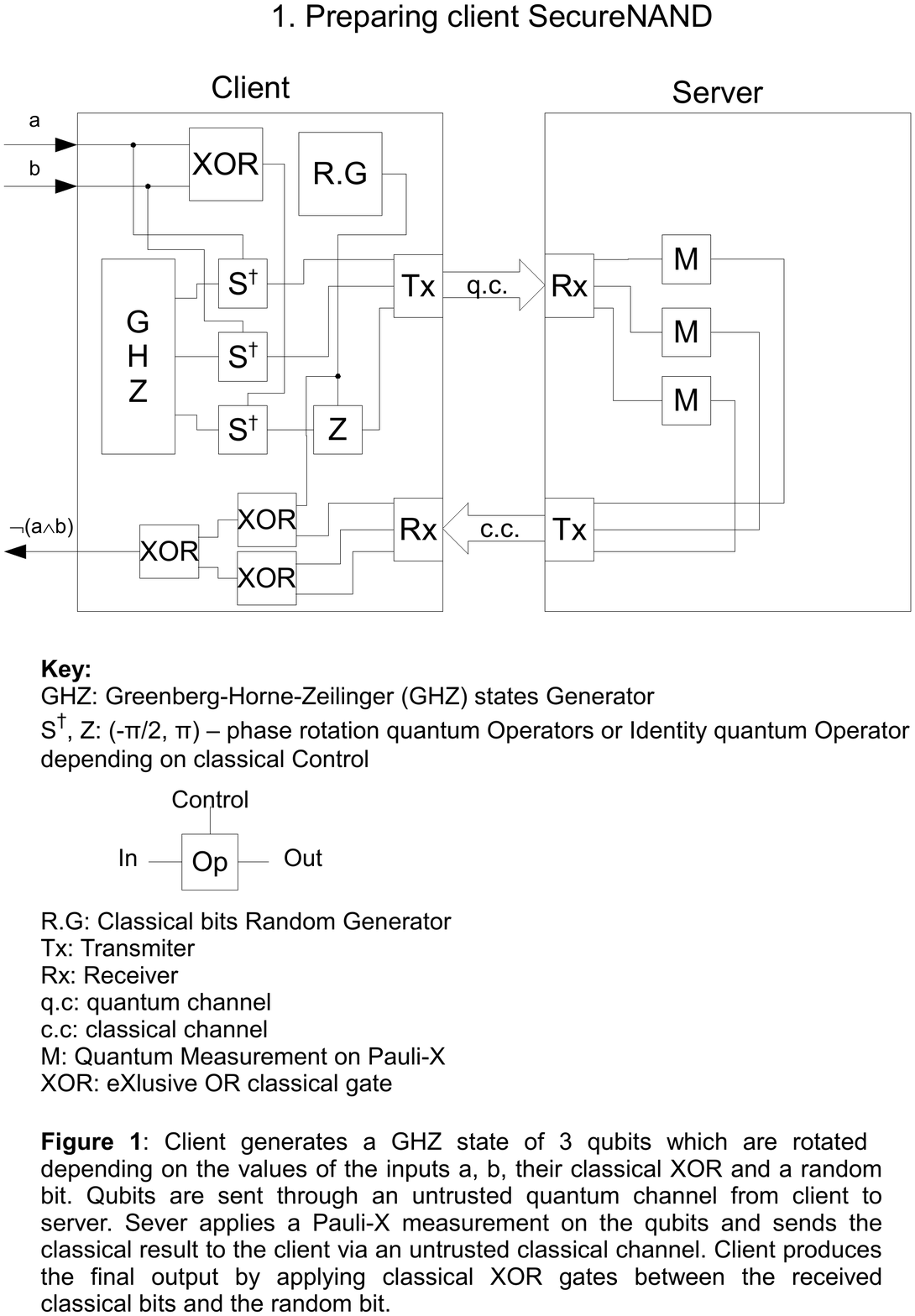} 
   \label{Fig-Implement1}
\end{figure*}

In this protocol client generates a GHZ state of 3 qubits which are rotated depending on the values of the inputs $a,b$, $a \oplus b$ and a random bit $r$. Qubits are sent through an untrusted quantum channel from client to server who applies a Pauli-$X$ measurement on the qubits and sends the classical result to the client via an untrusted classical channel. Client produces the final output by applying classical XOR gates between the received classical bits and the random bit (see Figure 1). In what follows we denote a random selection of an element of a set by $\in_{\R}$. 

\begin{algorithm}
\caption{Entangled-based Preparing Client SecureNAND}
 \label{prot:SecureNAND}
\begin{itemize}

\item Input (to Client): two bits $a,b$
\item Output (from Client): $\neg (a \wedge b)$
\item The Protocol:
\begin{itemize}

\item Client's round
\begin{enumerate}
\setlength{\itemsep}{0mm}
\item   $r\in_{\R} \{0,1\}$
\item Client generates 
\AR{
\ket{\Psi'} = Z_1^{r}{\left(S_1^\dagger\right)}^a 
{\left(S_2^\dagger\right)}^b  {\left(S_3^\dagger\right)}^{a \oplus b} \ket{\Psi}
}
and sends it to the Server.
\end{enumerate}
\item Server's round
\begin{enumerate}

\item Server measures the quits 1,2 and 3, with respect to the observables $X_1, X_2,$ and $X_3$, obtaining outcomes $b_1, b_2$ and $b_3$, respectively.
\item Server sends $b_1, b_2, b_3$ to Client
\end{enumerate}
\item Client's round
 \begin{enumerate}

 \item Client computes \EQ{out = b_1 \oplus  b_2 \oplus  b_3 \oplus r \label{SecureNAND:decode} }
 \item Client outputs $out$.
\end{enumerate}
\end{itemize}
\end{itemize} 
\end{algorithm}

We will say any SecureNAND protocol is \emph{correct} if for every run of the protocol where both players are honest (adhere to the protocol) and for all inputs $a,b$ we have 
\AR{
out =  \neg (a \wedge b) = 1 \oplus ( ab)
}
This definition will be used for all the presented protocols in this paper.
{Throughout this paper we will be using the notation for the logical \emph{and} between two bits $a,b$ as $a \wedge b$ and $ab$ interchangeably. }

\LE
Protocol \ref{prot:SecureNAND} is correct.
\EL
\proof First note that the protocol is correct if the following equality is true for all binary variables $a,b,r$:

\EQ{
X_1 X_2 X_3\ket{\Psi'} = (-1)^{1 \oplus ab \oplus  r} \ket{\Psi'}. \label{crit:core}
}
as this equality guarantees that the parity of the outcomes of measurements of Server equals $1 \oplus ab \oplus  r$ which implies Client will decode the correct outcome in Equation \ref{SecureNAND:decode} of Protocol \ref{prot:SecureNAND}.

In the remainder of the proof we define 
\AR{
P^{b} = \left\lbrace{ X,\ if\ b=0 \atop Y,\ if\ b=1  } \right.
}
and use the following Pauli and Clifford operator commutation relations:
\AR{
P^b Z^r = (-1)^{r} Z^r P^{b},\ \forall\ b,r \in \{0,1\}\\
P^{b} S^r = (-1)^{(b\oplus 1)r} S^r P^{b\oplus r},\ \forall\ b,r \in \{0,1\}\\
P^{b} \left(S^\dagger \right)^r = (-1)^{br}   \left(S^\dagger \right)^r P^{b\oplus r},\ \forall\ b,r \in \{0,1\}\\
}
and in particular the result that 
\AR{
X \left(S^\dagger \right)^r =    \left(S^\dagger \right)^r P^{r},\ \forall\ r \in \{0,1\}
}
and the result from \cite{BrowneGHZ} stating that 
\AR{
P_1^{a} P_2^{b} P_3^{a \oplus b} \ket{\Psi} = (-1)^{(1\oplus ab)}  \ket{\Psi},\ \forall\ a,b \in \{0,1\}.
}

We proceed to show the Equation (\ref{crit:core}) holds:
\AR{
X_1 X_2 X_3 \ket{\Psi'} = 
X_1 X_2 X_3  Z_1^{r}  {\left(S_1^\dagger\right)}^a {\left(S_2^\dagger\right)}^b  {\left(S_3^\dagger\right)}^{a \oplus b}  \ket{\Psi}=\\
 \left[X_1 Z_1^{r}{\left(S_1^\dagger\right)}^a \right]_1
 \left[ X_2 {\left(S_2^\dagger\right)}^b  \right]_2 \left[ X_3  {\left(S_3^\dagger\right)}^{a \oplus b} \right]_3 \ket{\Psi}=\\
 (-1)^r \left[Z_1^{r} {\left(S_1^\dagger\right)}^a P_1^{a}  \right]_1
 \left[ {\left(S_2^\dagger\right)}^b P_2^{b}   \right]_2 \left[  {\left(S_3^\dagger\right)}^{a \oplus b} P_3^{a\oplus b}  \right]_3 \ket{\Psi}=\\
 (-1)^r Z_1^{r} {\left(S_1^\dagger\right)}^a 
 {\left(S_2^\dagger\right)}^b  {\left(S_3^\dagger\right)}^{a \oplus b}  P_1^{a}  P_2^{b} P_3^{a\oplus b}   \ket{\Psi}=\\
  (-1)^{1\oplus ab \oplus r} Z_1^{r} {\left(S_1^\dagger\right)}^a 
 {\left(S_2^\dagger\right)}^b  {\left(S_3^\dagger\right)}^{a \oplus b}   \ket{\Psi}=(-1)^{1\oplus ab \oplus r}  \ket{\Psi'}
 }
In the derivation above we have simply used the trivial commutativity of operators acting on disjoint subsystems.
So the Lemma holds. \qed

\subsubsection{Security}

The desired security properties for out two-party protocols are the ability of hiding the secret information of the client (inputs bits) from the server, given formally below. This concepts is also refereed to as the blindness from the server's point of view. 

\DE
We will say any SecureNAND protocol is secure (also referred as \emph{blind}) if the cumulative state sent from Client to Server (averaged over Client's internal secret parameter $r$) is fixed (independent from the input $a$ and $b$). Again the same definition will be used for all other protocols. 
\ED

In other words, the system Server receives from Client could have been generated by Server without receiving any information from Client. In the remainder of this paper we will use the following short-hand: $$\bb{X} \df \op{X}{X},$$
for all labels $X$.

\LE
Protocol \ref{prot:SecureNAND} is blind.
\EL
\proof For fixed input $a,b$ the state Server receives from Client can be written as:
\EQ{
\sum_{r} \dfrac{1}{2} Z_1^{r} \eta   Z_1^{r} \label{all}
}
with 
\AR{
\eta =  {\left(S_1^\dagger\right)}^a  
{\left(S_2^\dagger\right)}^b  {\left(S_3^\dagger\right)}^{a \oplus b}  \ket{\Psi} \bra{\Psi} {\left(S_1\right)}^a 
{\left(S_2\right)}^b  {\left(S_3\right)}^{a \oplus b}  
}
Note that $\eta$ can be written as:
$\mathbf{S} \ket{\Psi} \bra{\Psi} \mathbf{S}^\dagger,$ where $$\mathbf{S} =  {\left(S_1^\dagger\right)}^a  
{\left(S_2^\dagger\right)}^b  {\left(S_3^\dagger\right)}^{a \oplus b}  $$
The operator $\mathbf{S}$ does not depend on the $r_i$ variables, and is diagonal in the computational basis so it commutes with Pauli Z operators. Using this commutation, the expression (\ref{all}) resolves as:
\AR{
\mathbf{S}\left(\sum_{r} \dfrac{1}{2} Z_1^{r} \ket{\Psi} \bra{\Psi}   Z_1^{r} \right)\mathbf{S}^\dagger
}
The operator $ \sum_{r}  \dfrac{1}{2} Z_1^{r} \ket{\Psi} \bra{\Psi}   Z_1^{r} $
can explicitly be written as 
\AR{
\sum_{r}  \dfrac{1}{2} Z_1^{r} \ket{\Psi} \bra{\Psi}   Z_1^{r} = \dfrac{1}{2} \left( \dfrac{1}{2}  \left( \dm{001} + \dm{110} - \ket{001}\bra{110} - \ket{110}\bra{001}\right) + \right.\\
\left. \dfrac{1}{2}  \left( \dm{001} + \dm{110} +\ket{001}\bra{110} + \ket{110}\bra{001}\right)\right) = \dfrac{1}{2} \left(
\bb{001} + \bb{110} \right)
}

Thus the operator above is diagonal in the computational basis, and commutes with $\mathbf{S}$ so we get: 
\AR{
\mathbf{S}\left(\sum_{r} \dfrac{1}{2} Z_1^{r} \ket{\Psi} \bra{\Psi}   Z_1^{r} \right)\mathbf{S}^\dagger = \left(\sum_{r} \dfrac{1}{2} Z_1^{r} \ket{\Psi} \bra{\Psi}   Z_1^{r} \right) \mathbf{S} \mathbf{S}^\dagger =\dfrac{1}{2} \left(
\bb{001} + \bb{110} \right)
}
This state is independent from $a$ and $b$ 
and the lemma is proved. \qed

\subsection{Measuring Client}

\begin{figure*}[h!] 
   \centering
   \includegraphics[width=300pt]{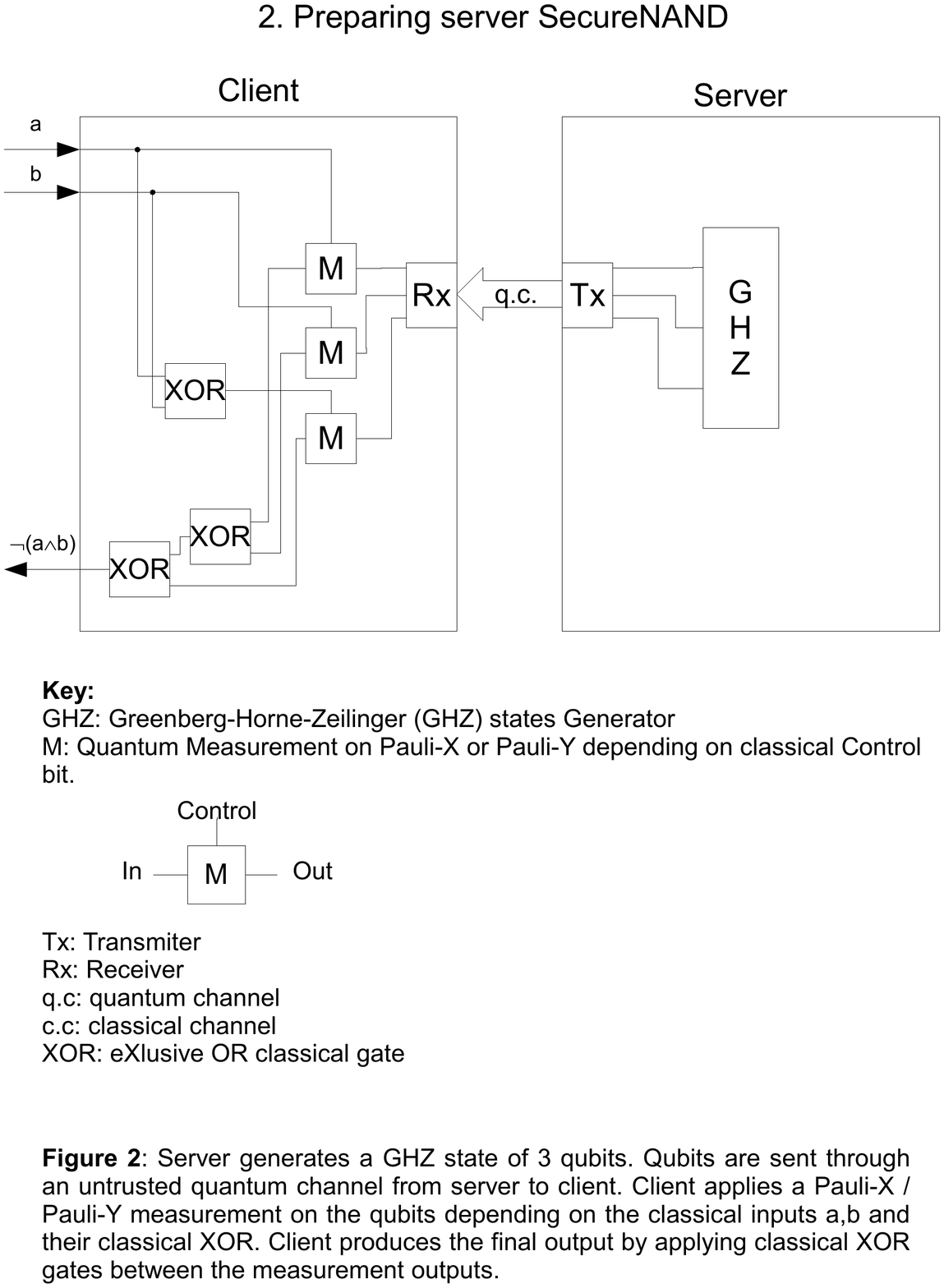} 
   \label{Fig-Implement2}
\end{figure*}

In this protocol server generates a GHZ state of 3 qubits. {The} qubits are sent through an untrusted quantum channel from {the} server to {the} client. {The} client applies a Pauli-$X$ or Pauli-$Y$ measurement on the qubits depending on the classical inputs $a$ and $b$ and their classical XOR. Client produces the final output by applying classical XOR gates between the measurement outputs (see Figure 2).

\begin{algorithm}[h!]
\caption{Entangled-based Measuring Client SecureNAND}
 \label{prot:SecureNANDMeasure}
\begin{itemize}

\item Input (to Client): two bits $a,b$
\item Output (from Client): $\neg (a \wedge b)$
\item The Protocol:
\begin{itemize}

\item Server's round
\begin{enumerate}
\setlength{\itemsep}{0mm}
\item The Server prepares the state $\ket{\Psi}$ and sends it to the Client
\end{enumerate}
\item Client's round
\begin{enumerate}

\item The client computes $c = a \oplus b$, measures the quits 1,2 and 3, with respect to the observables $P^{a}, P^{b},$ and $P^{c}$, obtaining outcomes $b_1, b_2$ and $b_3$, respectively.
 \item Client computes \EQ{out = b_1 \oplus  b_2 \oplus  b_3 \label{SecureNANDMeasure:decode} }
 \item Client outputs $out$.
\end{enumerate}
\end{itemize}
\end{itemize} 
\end{algorithm}

\LE
Protocol \ref{prot:SecureNANDMeasure} is blind and correct.
\EL
\proof The correctness of this protocol follows directly from the result in \cite{BrowneGHZ}. The blindness of the protocol trivially follows from the fact that no information is sent from the Client to the Server, thus the protocol is blind in all no signaling theories (including standard Quantum Mechanics).
\qed

\subsection{Bounce Protocol}

In this protocol we reduce the requirements on the client, which no longer has to measure or prepare states, but rather only modify states prepared by the server. Server generates a GHZ state of 3 qubits and sends them via an untrusted quantum channel to the client. Client applies {single-qubit} quantum operators depending on the values of the inputs $a, b$, $a \oplus b$ and 3 classical random bits. {The} client sends the rotated qubits to {the} sever via an untrusted quantum channel. {The} sever applies a Pauli-$X$ measurement on the qubits and sends the classical result to the client via an untrusted classical channel. The client produces the final output by applying classical XOR gates between the received classical bits and the random bits (see Figure 3).

\begin{figure*}[h!] 
   \centering
   \includegraphics[width=300pt]{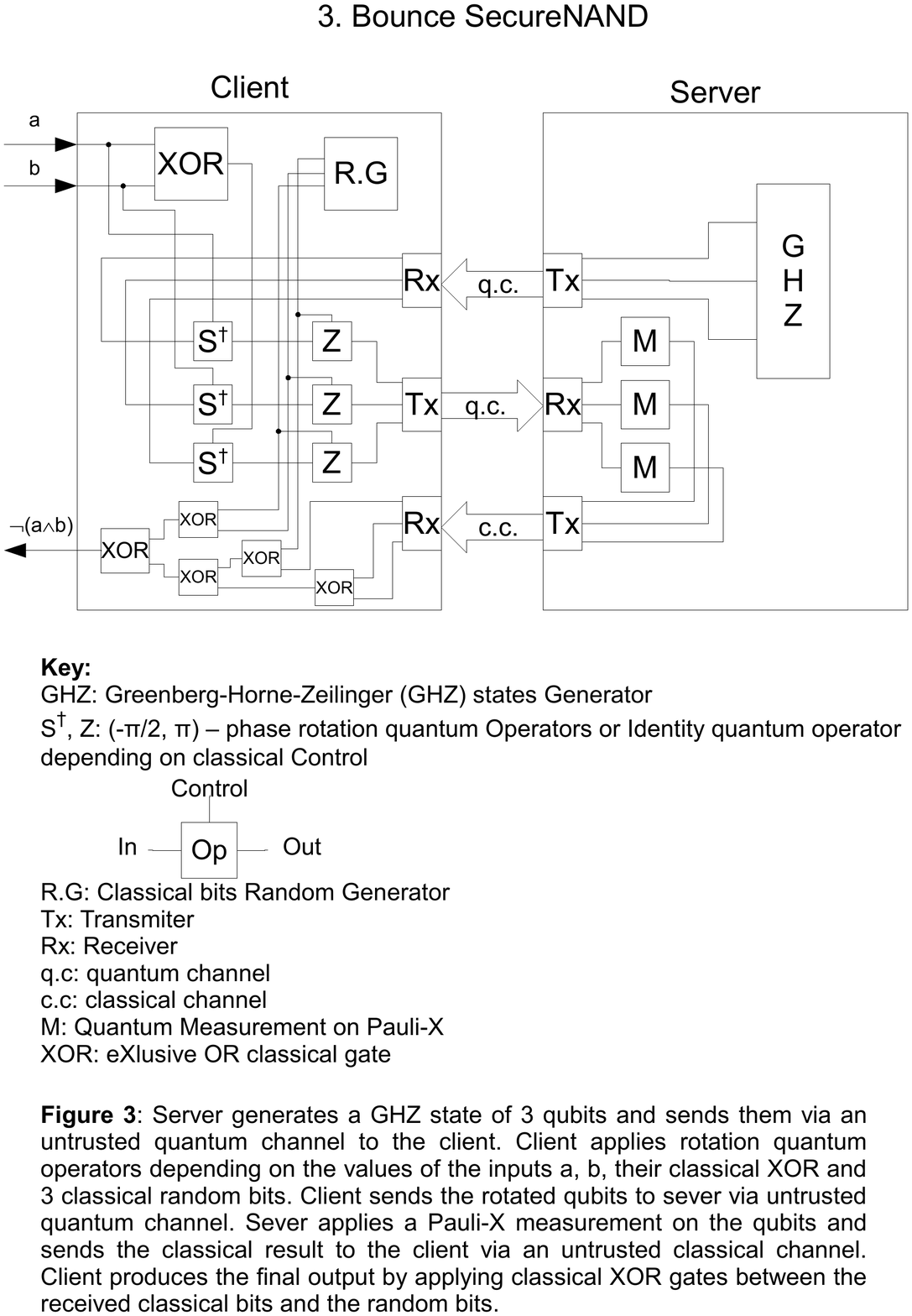} 
   \label{Fig-Implement3}
\end{figure*}

\begin{algorithm}
\caption{Entangled-based Bounce SecureNAND}
 \label{prot:BounceNAND}
\begin{itemize}

\item Input (to Client): two bits $a,b$
\item Output (from Client): $\neg (a \wedge b)$
\item The Protocol:
\begin{itemize}

\item Server's round
\begin{enumerate}
\setlength{\itemsep}{0mm}
\item The Server prepares the state $\ket{\Psi}$ and sends it to the Client
\end{enumerate}
\item Client's round
\begin{enumerate}
\setlength{\itemsep}{0mm}
\item Client receives the state $\ket{\Psi}$ from the server.
\item Client generates $r_1, r_2,r_3\in_{\R} \{0,1\}$
\item Client modifies the state $\ket{\Psi}$ to $\ket{\Psi'}$ as follows
\AR{
\ket{\Psi'} = Z_1^{r_1}Z_2^{r_2}Z_3^{r_3}{\left(S_1^\dagger\right)}^a 
{\left(S_2^\dagger\right)}^b  {\left(S_3^\dagger\right)}^{a \oplus b} \ket{\Psi}
}
and sends it to the Server.
\end{enumerate}
\item Server's round
\begin{enumerate}

\item Server measures the quits 1,2 and 3, with respect to the observables $X_1, X_2,$ and $X_3$, obtaining outcomes $b_1, b_2$ and $b_3$, respectively.
\item Server sends $b_1, b_2, b_3$ to Client
\end{enumerate}
\item Client's round
 \begin{enumerate}

 \item Client computes \EQ{out = b_1 \oplus  b_2 \oplus  b_3 \oplus r_1 \oplus r_2 \oplus r_3. \label{BounceNAND:decode} }
 \item Client outputs $out$.
\end{enumerate}
\end{itemize}
\end{itemize} 
\end{algorithm}

\LE
Protocol \ref{prot:BounceNAND} is correct.
\EL
\proof The correctness is directly obtained from the correctness of the Entangled-based Preparing Client SecureNAND. To see this note that the states the server performs the measurements on are identical in the two protocols, up to the existence of possible $Z_2^{r_2}$ and $Z_3^{r_3}$ rotations on the second and third qubit.
Since we both have that
\AR{
X Z^{r} = (-1)^r Z^{r} X, \textup{and}\\
Y Z^{r} = (-1)^r Z^{r} Y,
}
these rotations cause an additional (multiplicative) phase of $(-1)^{r_2 \oplus r_3}$.
But this is compensated for in the modified decoding of the client in stage \ref{BounceNAND:decode} so the output is correct in this protocol as well.
So the Lemma holds. \qed

\LE
Protocol \ref{prot:BounceNAND} is blind.
\EL
\proof For fixed input $a,b$ the state server obtains in the protocol can be written as:
\EQ{
\sum_{r_1, r_2, r_3} \dfrac{1}{8}\left( Z_1^{r_1} Z_2^{r_2} Z_3^{r_3}\otimes \mathbbmss{1}_S \right) \eta  \left( Z_1^{r_1}Z_3^{r_2}Z_3^{r_3}\otimes \mathbbmss{1}_S \right) \label{all}
}
with 
\AR{
\eta =  \left({\left(S_1^\dagger\right)}^a  
{\left(S_2^\dagger\right)}^b  {\left(S_3^\dagger\right)}^{a \oplus b} \otimes \mathbbmss{1}_S  \right)  \rho^{S}  \left({\left(S_1\right)}^a 
{\left(S_2\right)}^b  {\left(S_3\right)}^{a \oplus b}\otimes \mathbbmss{1}_S \right) , 
}
where  $\rho^{S}$ is any state the malevolent server could have {initially} prepared.
Note that the actions of the client are only on {a subsystem} of the {whole} system {in the state} $\rho^{S}$, signifying that the server might have prepared an entangled state, and sent only a {subsystem} to the client to be modified, while keeping the remainder of the {system}.

Since $Z$ operators commute with the phase $S^\dagger$ operators, and the parameters of the phase operators do not depend on $r_i$ values, by introducing the shorthand 
$\mathbf{S} = \left({\left(S_1^\dagger\right)}^a  
{\left(S_2^\dagger\right)}^b  {\left(S_3^\dagger\right)}^{a \oplus b} \otimes \mathbbmss{1}_S  \right) $
we can rewrite the state of the server's system as:

 \AR{
\left(\mathbf{S}\otimes \mathbbmss{1}_S  \right) \sum_{r_1, r_2, r_3} \dfrac{1}{8}\left( Z_1^{r_1} Z_2^{r_2} Z_3^{r_3}\otimes \mathbbmss{1}_S \right) \rho^S  \left( Z_1^{r_1}Z_3^{r_2}Z_3^{r_3}\otimes \mathbbmss{1}_S \right) \left(\mathbf{S}^\dagger\otimes \mathbbmss{1}_S \right).
}

The state $\rho^S$ has two partitions - the partition corresponding to the subsystem the server sends to the client, and the subsystem he keeps.
Thus $\rho^S$ can be written (in the Pauli operator basis) as:
\AR{
\sum_{i,j} \alpha_{i,j} \underbrace{\sigma_i}_{C} \otimes \underbrace{\sigma_j}_{S'}
}
where $C$ denotes the subsystem sent to the client, and $S'$  the subsystem kept by the server, and $\sigma_i$ and $\sigma_j$ denote general Pauli operators acting on the two respective subsystems.

Next, we have the following derivation:
\AR{
\sum_{r_1, r_2, r_3} \dfrac{1}{8}\left( Z_1^{r_1} Z_2^{r_2} Z_3^{r_3}\otimes \mathbbmss{1}_S \right) \rho^S  \left( Z_1^{r_1}Z_3^{r_2}Z_3^{r_3}\otimes \mathbbmss{1}_S \right) =\\
\sum_{r_1, r_2, r_3} \dfrac{1}{8}\left( Z_1^{r_1} Z_2^{r_2} Z_3^{r_3}\otimes \mathbbmss{1}_S \right) \sum_{i,j} \alpha_{i,j} \underbrace{\sigma_i}_{C} \otimes \underbrace{\sigma_j}_{S'} \left( Z_1^{r_1}Z_3^{r_2}Z_3^{r_3}\otimes \mathbbmss{1}_S \right) =\\
\dfrac{1}{8} \sum_{i,j}  \alpha_{i,j} \left(  \sum_{r_1, r_2, r_3} Z_1^{r_1} Z_2^{r_2} Z_3^{r_3}  \sigma_i  Z_1^{r_1}Z_3^{r_2}Z_3^{r_3} \right) \otimes  \sigma_j  =
}
Note that since both $X$ and $Y$ anticommute with $Z$, the expression 
$$ \sum_{r_1, r_2, r_3} Z_1^{r_1} Z_2^{r_2} Z_3^{r_3}  \sigma_i  Z_1^{r_1}Z_3^{r_2}Z_3^{r_3} $$ is non-zero only if all the single qubit operators making up $\sigma_i$ are either $Z$ or identity, and in both cases diagonal in the computational basis.
Thus, we can write the final expression of the derivation above as:
\AR{
 \sum_{i,j}  \alpha'_{i,j}   \sigma'_i   \otimes  \sigma_j 
}
where $\sigma'_i $ is diagonal in the computational basis.

So, overall, for the state of the server's system we have:
\AR{
\left(\mathbf{S}\otimes \mathbbmss{1}_S  \right) \sum_{r_1, r_2, r_3} \dfrac{1}{8}\left( Z_1^{r_1} Z_2^{r_2} Z_3^{r_3}\otimes \mathbbmss{1}_S \right) \rho^S  \left( Z_1^{r_1}Z_3^{r_2}Z_3^{r_3}\otimes \mathbbmss{1}_S \right) \left(\mathbf{S}^\dagger\otimes \mathbbmss{1}_S \right) = \\
\left(\mathbf{S}\otimes \mathbbmss{1}_S  \right)  \sum_{i,j}  \alpha'_{i,j}   \sigma'_i   \otimes  \sigma_j   \left(\mathbf{S}^\dagger\otimes \mathbbmss{1}_S \right) =\\
  \sum_{i,j}  \alpha'_{i,j}   \left(\mathbf{S}\otimes \mathbbmss{1}_S  \right)  \sigma'_i   \otimes  \sigma_j   \left(\mathbf{S}^\dagger\otimes \mathbbmss{1}_S \right)
}
and since $\sigma'_i$ commute with $\mathbf{S}$ we get:
\AR{
 \sum_{i,j}  \alpha'_{i,j}    \sigma'_i  \mathbf{S} \mathbf{S}^\dagger  \otimes  \sigma_j  =
  \sum_{i,j}  \alpha'_{i,j}    \sigma'_i   \otimes  \sigma_j 
}
Since $\alpha'_{i,j}$ is independent from $a$ and $b$, this state is independent from $a$ and $b$ 
and the lemma is proved. \qed

\begin{algorithm} [h!] 
\caption{Single Qubit Bounce SecureNAND}
 \label{prot:BounceNAND1Q}
\begin{itemize}

\item Input (to Client): two bits $a,b$
\item Output (from Client): $\neg (a \wedge b)$
\item The Protocol:
\begin{itemize}

\item Server's round
\begin{enumerate}
\setlength{\itemsep}{0mm}
\item The Server prepares the state $\ket{+}$ and sends it to the Client
\end{enumerate}
\item Client's round
\begin{enumerate}
\setlength{\itemsep}{0mm}
\item Client receives the state $\ket{+}$ from the server.
\item Client generates $r\in_{\R} \{0,1\}$
\item Client modifies the state $\ket{+}$ to $\ket{\Psi}$ as follows
\AR{
\ket{\Psi} = Z^{r} {S}^a 
{S}^b  {\left(S^\dagger\right)}^{a \oplus b} \ket{+}
}
and sends it to the Server.
\end{enumerate}
\item Server's round
\begin{enumerate}

\item The server measures the qubit with respect to the $X$ basis, obtaining the outcome $s$
\item Server sends $s$ to Client
\end{enumerate}
\item Client's round
 \begin{enumerate}

 \item Client computes \EQ{out = s \oplus r\oplus 1 \label{BounceNAND1Q:decode} }
 \item Client outputs $out$.
\end{enumerate}
\end{itemize}
\end{itemize} 
\end{algorithm}

\section{Single Qubit Protocols}

Here, we give variants of a new class of secure NAND protocols which only require single qubit manipulations. Similarly to the variants we have given for the GHZ-based protocols, the single qubit protocol can also be modified to a client preparation or a measuring client protocol. In the former, it is the client which would prepare the initial $\ket{+}$ state, whereas in the measuring client protocol, the client would perform the final measurements. Both protocols are blind and correct as a simple consequence of the Single Qubit Bounce SecureNAND protocol (that we describe below) see Figure 4, 5 and 6 in Appendix A.  
{Thus, here we} only {need to} analyse the Single Qubit Bounce SecureNAND protocol (Protocol \ref{prot:BounceNAND1Q}). 

{The} server generates a single qubit state and sends it via an untrusted quantum channel to the client. Client applies a series of rotation quantum operators depending on the values of the inputs $a$, $b$, $a \oplus b$, and a classical random bit. Client sends the rotated qubit to sever via untrusted quantum channel. Sever applies a Pauli-$X$ measurement on the qubit and sends the classical result to the client via an untrusted classical channel. Client produces the final output by applying classical XOR gates between the received classical bit, a classical bit in state 1 and the random bit (Figure 6 in Appendix A). To see the correctness note that if the server was honest, it is a straightforward calculation to see the state of the qubit the server receives is
\AR{
Z^{r} Z^{a\wedge b} \ket{+}
}
Then the result of the measurement performed by the server is $s = r \oplus a\wedge b$, and the decoding produces $out = 1 \oplus a \wedge b$ as required.

To see the security, note that the most general strategy of the server is to prepare a bipartite state $\pi_{1,2}$ and send the first subsystem to the client. Then the state of the server system (up to a normalization factor $1/2$), once the client performed her round is:
\AR{
\sum_{r} \left( Z^{r} Z^{a\wedge b} \otimes \mathbbmss{1}_2 \right) \pi_{1,2}  \left( Z^{r} Z^{a\wedge b} \otimes \mathbbmss{1}_2 \right) = 
\sum_{r'} \left( Z^{r'}  \otimes \mathbbmss{1}_2 \right) \pi_{1,2}  \left( Z^{r'} \otimes \mathbbmss{1}_2 \right)
}
where $r' = r \oplus a\wedge b$. Since $r$ is distributed uniformly at random, so is $r'$ so the state above does not depend on $a$ or $b$.

\section{Impossibility Results}

The main result of this section is to prove the optimality of our protocol. We first prove that it is impossible to achieve the similar task of secure delegated computing of our protocols by removing the quantum requirement. Next we show that in the quantum case, the quantum states must depend on the input of the client as it is done in our protocols. This result also indicates that a quantum off-line protocol could not be achieved.
\TH
No classical protocol, in which the client is restricted to XOR computations can delegate deterministically computation of NAND to a server while keeping the blindness.
\HT
\proof We prove this result first for the case of two rounds of communication, and no initial shared randomness.
Any such protocol will have the following three stages: client's encoding, server's computation, and client's decoding.

\noindent\emph{Client's encoding.}
In this stage, the only thing the client can do is to compute $C_1(a,b,\ora{x})$, where $a,b$ are the input bits,
$\ora{x}$ is a random bit string (of any length) and $C_1$ is a computation which can be implemented using only XOR gates.
However, the state $C_1(a,b,\overrightarrow{x})$ must be independent from $a$ and $b$ to maintain blindness when averaged over all $\overrightarrow{x}$.

\noindent\emph{Server's computation.} The only thing the server can do is to apply some computable function $S$ on $C_1(a,b,\overrightarrow{x})$, thus returning
$S(C_1(a,b,\overrightarrow{x}))$.

\noindent\emph{Client's decoding.} The only thing the client can do is to run some function $C_2$, on all the data he has, which is implementable using XOR gates only:
\AR{
C_2(a,b,\overrightarrow{x},S(C_1(a,b,\overrightarrow{x}))) = NAND(a,b)\ \textup{(correctness)}
}
and the output must (deterministically) be the NAND of the inputs.

Let $c = C_1(a',b',\overrightarrow{x'})$ be some constant the client may send to the server.
Then, because of blindness it must hold that for all $a,b$ there must exist $\overrightarrow{x}(a,b)$, which depends on $a,b$ such that
\AR{
C_1(a,b,\overrightarrow{x}(a,b))=c.
}
To see this, note that if the client could send $c$, but not for some inputs $a''$ and $b''$, then upon receiving $c$ the server learns something about the input, namely that it is not $a'',b''$, which violates blindness.
Note also that since all the computations the client can perform use only XOR gates (and without the loss of generality, reversible), the client \emph{can} compute $\overrightarrow{x}(a,b)$ given $a,b$ {using only XOR operations}. But then, by the correctness of the protocol we have that
$$
C_2(a,b,\overrightarrow{x}(a,b),S(c) ) = NAND(a,b)\ \textup{(correctness)}.
$$
But $S(c)$ is constant as well.
This implies that given a fixed string $S(c)$ the client can compute the NAND of any input using just XOR gates, which is not possible.

This argument {can be further} generalized to a setting with shared randomness and many rounds of communication.
It is easy to see that the randomness cannot help as the protocol must be deterministic (hence work for any sampling of the joint random variable), whereas using multiple rounds (all of which must be independent of the input, {from the viewpoint of the server}) just yields a longer constant string (analogous to $S(c)$) using which the client can compute the $NAND$ on her own, which is again impossible.
\qed

The above result highlights the magic of quantum phase exploited in our protocol where despite sending a quantum state dependent on the input, {the input bits} remain inaccessible to the server. 
{Next}, following a similar line of argument we prove that one cannot hope for an improvement of our protocols i.e. a {quantum} offline procedure (similar to universal blind quantum computing \cite{UBQC}) where the initial quantum states communicated in the protocol {are} independent of the secret input.  

\subsection{Impossibility of Offline Communication}

We begin by addressing protocols with two rounds of communication between the client and the server. By round we refer to an instance of either the client sending a message to the server, or the server sending {a message} to the client. Since the last message, for it to have any meaning, must come from the server, the order of the two rounds is client $\rightarrow$ server, followed by server $\rightarrow$ client. The generic description (definition) of a potential secure NAND quantum offline protocol with two rounds is given later in Protocol 7. In order to prove the impossibility of obtaining such a protocol we prove several lemmas proving first the impossibility of a particular class of somehow `minimal' NAND quantum offline protocols (see Protocol 5 and 6 below). {Following this,} we present the reduction between these protocols i.e. if a generic protocol of type Protocol 7 is possible then so is the minimal protocol, hence proving the impossibility of obtaining any offline quantum protocol.

These types of protocols are intimately linked to the composability of secure NAND computations in a larger computation {\footnote{{The security issues of composability of our protocols we do not explicitly address in this paper. However, we do note that the lower bounds on what is possible we establish here imply that the impossibility results will also hold true in any composable security setting.}}}. Note that since, for the second layer of any computation, the client does not know the inputs in advance (since he cannot compute them herself) but knows the encryption of the outputs in advance, thus,  quantum offline protocols are necessary and probably sufficient for the composition of NANDs in a larger computation, without requiring additional run-time communication. The case where run-time communication is allowed will be studied presently. Note also that it does not matter what function, {which in tandem with XOR and NOT gates forms a universal set}, we use. For simplicity, here we focus on AND. 

The simple quantum offline secure AND computation with two rounds of communication (Simple AND QO2, Protocol 5) is the most natural first attempt, which is inspired by information-theoretic considerations - since the client's input is two bits $a$ and $b$, hence the quantum state encodes two bits of $x$ and $y$. Therefore to hide the two bits in the quantum state, additional randomness of two bits $r_1$ and $r_2$ is needed. 

 \begin{algorithm}[h!]
\caption{Simple SecureAND QO2}
 \label{prot:SimpleNAND}
The functionality of the Small AND protocol:
\begin{itemize}

\item Input (to the client): two bits $a,b$
\item Output (from the client): $(a \wedge b)$
\item The Protocol:
\begin{itemize}

\item Client's round
\begin{enumerate}

\item Client generates a quantum state $\rho^{x,y}_{r_1, r_2}$, characterized by random bits $x,y,r_1,r_2$ and sends it to the server.

\item Client receives her input bits $a,b$.
\item Client computes $m_c = (x\oplus a, y \oplus b)$ and sends it to the server. 
 \end{enumerate}
\item Cerver's round
\begin{enumerate}

\item Server performs a (generalized) measurement of $\rho^{x,y}_{r_1,r_2}$, parametrized by $m_c$. He obtains the outcome $m_s$ and sends it to the client.
\end{enumerate}
\item Client's round
 \begin{enumerate}

 \item Client computes $out = m_s \oplus r_1 \oplus r_2$.
 \item Client outputs $out$.
\end{enumerate}
\end{itemize}
\end{itemize} 
\end{algorithm}

To shorten our expressions, in this section we will be predominantly use $ab$ to denote the logical \emph{and} of two bits $a,b$.

Recall that the correctness of these protocols are defined by requesting $out = ab$, and blindness is defined by the equation 
\AR{
\sum_{\mathbf{x}}  \bb{m(a,b)} \otimes \rho^{\mathbf{x}} = \eta \;\; \forall a,b,
} 
{where $a,b$ are the input bits, $m(a,b)$ the classical message which may depend on the input, $\rho^x$ a quantum state which depends on some random parameters $x$ (but may also depend on $a,b$), and $\eta$ is a positive-semidefinite operator, independent from $a,b$} \footnote{We are omitting any normalization factors, so $\eta$ may be of non-unit trace.}.

\LE
No Simple SecureAND QO2 can be correct and blind.
\EL
\proof
As in any Simple SecureAND QO2 protocol the client sends two classical bits of information to the server (here denoted $a',b'$), without the loss of generality, we may assume that the message the server returns to the client is a single bit measurement outcome of one of four (generalized) measurements (one for each message $(a',b')$) which we denote
$M^{a',b'}(\rho^{x,y}_{r_1,r_2}).$
The correctness of the protocol entails that
\AR{M^{a',b'}(\rho^{x,y}_{r_1,r_2}) = (a'\oplus x)(b' \oplus y) \oplus r_1\oplus r_2}
For clarity we briefly comment on the equation above. Since, for message $(a',b')$ the server performs a generalized two-outcome measurement, this measurement can be represented by the POVM elements $\Pi^{a',b'}_0, \Pi^{a',b'}_1$ (which are positive operators summing to the identity), corresponding to outcomes 0 and 1, respectively. Then the equation above means that
\AR{
Tr( \Pi^{a',b'}_{(a'\oplus x)(b' \oplus y) \oplus r_1\oplus r_2} \rho^{x,y}_{r_1,r_2} ) =1
}
Then, by taking $r=r_1 \oplus r_2$ and defining $\rho_r^{x,y} = 1/2 (\rho^{x,y}_{0,r} + \rho^{x,y}_{1,1\oplus r})$ we get, by linearity, that 
\AR{M^{a',b'}(\rho^{x,y}_{r}) = (a'\oplus x)(b' \oplus y) \oplus r,} or equivalently, 
\AR{
Tr( \Pi^{a',b'}_{(a'\oplus x)(b' \oplus y) \oplus r} \rho^{x,y}_{r} ) =1
} and also that 
\AR{ Tr( \Pi^{a',b'}_{(a'\oplus x)(b' \oplus y) \oplus r} \rho^{x,y}_{r \oplus 1} ) =0}
The two equations above immediately entail that 
$ \rho^{x,y}_{r}$ and $ \rho^{x,y}_{r\oplus 1}$ must be (mixtures of mutually) orthogonal states, which we denote $ \rho^{x,y}_{r} \bot \rho^{x,y}_{r\oplus 1}.$
But, more generally, the equations above imply that two states $\rho^{x,y}_r$ and $\rho^{x',y'}_{r'}$ must be in orthogonal subspaces, whenever any of the sub/superscripts differ. To see this, we will consider the remaining cases separately.
First, assume that $r=r',$ but $x\not= x'$ and/or $y\not= y'$.
Then if we set $a'=x\oplus 1$ and $b' = y \oplus 1$ we see that 
\AR{
M^{a',b'}(\rho^{x,y}_{r}) = (a'\oplus x)(b' \oplus y) \oplus r = 1\oplus r
} 
but
\AR{
M^{a',b'}(\rho^{x',y'}_{r}) = (a'\oplus x')(b' \oplus y') \oplus r = r
} so the outcomes \emph{deterministically} differ, meaning that the two states must be in orthogonal subspaces.
We have already seen that the same conclusion follows if  $r\not = r',$ and $x = x'$ and $y = y'$.
The next case is when $r\not = r',$ and either  $x\not= x'$ or $y\not= y'$ (but one is an equality).
Assume that $x=x'$, $y\not = y'$ and $r=0$.
Then if we set $a'=x=x'$ we see that 
\AR{M^{x,b'}(\rho^{x,y}_{0}) = (x\oplus x)(b' \oplus y) = 0} and
\AR{M^{x,b'}(\rho^{x,y'}_{1}) = (x\oplus x')(b' \oplus y') \oplus 1 = 1.}
Similarly, if $r=1$ we get opposite results, and if $x\not=x'$ and $y=y'$ we get the same by setting $b'=y=y'$.
Finally, we must consider the case when all the parameters differ.
First, assume $r=0$, then by setting $a'=x$ and $b'=1\oplus y$ we get:
\AR{M^{a',b'}(\rho^{x,y}_{0}) = (x\oplus x)(b' \oplus y) = 0,\  \textup{and}\\
M^{a',b'}(\rho^{x',y'}_{1}) = (x\oplus x')(1\oplus y \oplus y') \oplus 1 = 1} since $y\not= y'$, if $r=1$ then the first equation above would yield 1, and the last would yield 0, since $1 \oplus y \oplus y'=0$. Thus we can conclude that the states $\{ \rho_r^{x,y} \}_{x,y,r}$ are all in orthogonal subspaces. But this means, in particular, that the states $1/4 (\sum_{r_1,r_2} \rho^{x,y}_{r_1,r_2})$ are in orthogonal subspaces for all $x,y$ which implies that there exists a measurement which perfectly reveals $x$ and $y$ given any  $\rho^{x,y}_{r_1,r_2}$.
Thus, the server can perfectly learn $x$ and $y$ and, given the classical message of the client, the inputs of the client, and the protocol is not blind. 
\qed 

In the above proof we have quickly concluded that the two bits $r_1,r_2$ are superfluous and one will suffice ({which is intuitive as only one random bit is needed to one-time pad the one bit outcome}). This gives us the definition of the next general family of protocols (Small AND QO2, Protocol 6) as we describe below and will refer to later. 
 
\begin{algorithm}[h!]
\caption{Small SecureAND QO2}
 \label{prot:SmallNAND}
The functionality of the Small AND protocol:
\begin{itemize}

\item Input (to the client): two bits $a,b$
\item Output (from the client): $(a \wedge b)$
\item The Protocol:
\begin{itemize}

\item Client's round
\begin{enumerate}

\item Client generates a quantum state $\rho^{x,y}_{r}$, characterized by random bits $x,y,r$ and sends it to the server.

\item Client receives her input bits $a,b$.
\item Client computes $m_c = (x\oplus a, y \oplus b)$ and sends it to the server. 
 \end{enumerate}
\item Server's round
\begin{enumerate}

\item Server performs a (generalized) measurement of $\rho^{x,y}_{r}$, parametrized by $m_c$. He obtains the outcome $m_s$ and sends it to the client 
\end{enumerate}
\item Client's round
 \begin{enumerate}

 \item Client computes $out = m_s \oplus r$.
 \item Client outputs $out$.
\end{enumerate}
\end{itemize}
\end{itemize} 
\end{algorithm}

\LE
No small SecureAND QO2 can be correct and blind.
\EL
 \proof
 Obvious from the proof of impossibility of simple AND QO2, where we have actually reduced simple to small protocols. \qed

\subsection{Generalization: QO2}

\begin{algorithm}[h!]
\caption{SecureAND QO2}
 \label{prot:NAND}
The functionality of the AND protocol:
\begin{itemize}

\item Input (to the client): two bits $a,b$
\item Output (from the client): $(a \wedge b)$
\item The Protocol:
\begin{itemize}

\item Client's round
\begin{enumerate}

\item Client generates a quantum state $\rho^{\mathbf{x}}$, characterised by a sequence of random parameters $\mathbf{x} = (x_1, \ldots, x_n),$ and sends it to the server.

\item Client receives her input bits $a,b$ (the client could have had her bits all along. It is however the defining property of quantum-offline protocols that the parameters $\mathbf{x}$ are independent from $a,b$).

\item Client computes an XOR-computable function 
\AR{
m_c = \text{XOR}_E (a,b, \mathbf{x})
} 
(E for encryption) of the input and the random parameters. Note that it would be superfluous for the client to generate additional random values at this stage - they could be part of $\mathbf{x}$, without influencing the state the client generates.
\item Client sends $m_c$ to the server.
 \end{enumerate}
\item Server's round
\begin{enumerate}

\item Server performs a (generalized) measurement of $\rho^{\mathbf{x}}$, parametrized by $m_c$. He obtains the outcome $m_s$ and sends it to the client.
\end{enumerate}
\item Client's round
 \begin{enumerate}
\item Client computes an XOR-computable function 
\AR{
out = \text{XOR}_D(a,b,\mathbf{x},m_s)
} 
(D for decryption).
 \item Client outputs $out$.
\end{enumerate}
\end{itemize}
\end{itemize} 
\end{algorithm}

In order to prove a reduction between the general case of Protocol 7 and the simple scenario of Protocol 6 we start with a supposedly given blind and correct QO2 protocol and iteratively construct a blind correct small QO2, using a sequence of claims which define increasingly simpler protocols.

\TH \label{t-qoffline}
If there exists a blind, correct SecureAND QO2 then there exists a  blind correct \emph{Small} SecureAND QO2.
\HT

The objects which appear in the protocol (which differ from the objects in the small QO2) are as follows:
\AR{
\rho^{\mathbf{x}},\; \text{with}\ \mathbf{x} = (x_1, \ldots, x_n)\ -\ \textup{the\ quantum\ state\ parametrized\ by\ } n\ \textup{bits}\\
m_c = \text{XOR}_E (a,b, \mathbf{x})\ - \ \textup{the\ m\ bit\ message\ from\ the client}\\
m_s, \; \ \text{the\ k\ bit\ message\ from\ the server}\\
ab = out = \text{XOR}_D(a,b,\mathbf{x},m_s)\ - \ \textup{the\ calculation\ of\ the\ output}\\
}\\

\noindent {\bf Claim 1.} Nothing is gained from using multi-bit $m_s$.
\proof
Note that since the client is restricted to computing XOR operations, we can dissect 
\AR{
\text{XOR}_D(a,b,\mathbf{x},m_s)
} 
and see that it must be of the form
\AR{
\text{XOR}_D(a,b,\mathbf{x},m_s) = \text{XOR}'_D(a,b,\mathbf{x}) \oplus \bigoplus_{j \in I \subseteq \left[k \right]} [m_s]_j,
}
{where $[m_s]_j$ is the $j^{th}$ bit of the $k$-bit message $m_s$}.
That is, it is a mod 2 addition of something which does not depend on the server's message, and the mod 2 addition of some of the bits of the message responded by the server. 
Since the form of the message (\emph{i.e} the explicitly description of the function $\text{XOR}_D$) is public, being in the protocol description, the protocol remains secure and correct if the server himself computes the bit {$\bigoplus_{j \in I \subseteq \left[k \right]}[m_s]_j$}, and returns this to the client. Thus, for every correct, blind QO2 there exists a correct blind QO2$_1$ where the server's message comprises only one bit.
The remainder of the claims assumes we are dealing with a QO2$_1$ protocol.\qed

\noindent {\bf Claim 2.} No random parameters which do not appear in the encryption or decryption are needed.
\proof
Let $S \subset \left[ n \right]$ be a subset of indices of the random parameters which appear in either encryption (as variables of $\text{XOR}_E$) or decryption ($\text{XOR}_D$), and let $S' = \left[ n \right] \setminus S$ be the subset which does not appear.
Then, by exchanging the state $\rho^\mathbf{x}$ with the state  $$(\rho')^\mathbf{x'} = \sum_{x_j \vert j\in S'} \dfrac{1}{2^{\vert S' \vert}} \rho^\mathbf{x}$$ in a QO2$_1$ protocol it is easy to see we again obtain a protocol ({which we refer to as} QO2$_2$) which is correct and blind.  In QO2$_2$ protocols, all the random parameters appear either in the decryption or encryption.
The remainder of the claims assumes we are dealing with a QO2$_2$ protocol.\qed

\noindent {\bf Claim 3.} No more than one random parameter which appears only in the decryption is needed.
\proof
Let $S_{D\setminus E} \subset \left[ n \right]$ be the set of indices of random parameters which appear only in the decryption, that is, as a variable of the function $\text{XOR}_D$. Without the loss of generality, we will assume that the last $k$ indices are such.
Then $\text{XOR}_D(a,b,\mathbf{x},m_s)$ (due to the restrictions on the client) can be written as:

\AR{
\text{XOR}_D(a,b,\mathbf{x},m_s) = \text{XOR}'_D(a,b,m_s,x_1\ldots,x_{n-k}) \oplus x_{n-k+1} \oplus \cdots \oplus x_{n},
}
Then, by exchanging the state $\rho^\mathbf{x}$ with the state  
\AR{
(\rho')^{x_1, \ldots, x_{N-k},x} = \sum_{\substack{x_j \vert j\in S_{D\setminus E}\\ s.t. \\ \oplus_j x_j = x}} \dfrac{1}{2^{\vert S_{D\setminus E}  \vert -1}} \rho^\mathbf{x}} 
in a QO2$_2$ protocol we again obtain a protocol ({which we refer to as} QO2$_3$) which is correct and blind.  
Blindness is trivial, as the sum over all the random parameters for the state $\rho^{\mathbf{x}}$ yields the same density operator as the sum over all random parameters for the state $(\rho')^{x_1, \ldots, x_{n-k},x} $ (and no message correlated to the summed up random parameters is sent from the client to the server).
Correctness holds as the correctness of the (original) QO2$_2$ protocol only depended on the parity of the $k$ random parameters, and the construction above preserves this. \qed

In QO2$_3$ protocols, at most one random parameter appears in the decryption only.
The remainder of the claims assumes we are dealing with a QO2$_3$ protocol.

\noindent {\bf Claim 4.} Client's input bits $a$ and $b$ do not need to appear in the decryption function.
\proof
In general the decryption function of the client (for QO2$_3$) protocols attains the form
\AR{
\text{XOR}_D(a,b,\mathbf{x},m_s) = \text{XOR}'_D(a,b,m_s) \oplus \bigoplus_{j \in  S_{E \cap D}} x_j \oplus x_n \ or \\
\text{XOR}_D(a,b,\mathbf{x},m_s) = \text{XOR}'_D(a,b,m_s) \oplus \bigoplus_{j \in  S_{E \cap D}} x_j
}
where $S_{E \cap D}$ is the set of indices of random parameters which appear in both the decryption and encryption function, and $x_n$ may appear only in the decryption function. Here, we have assumed without the loss of generality that it is the last random parameter that (possibly) appears only in the decryption function.
First, we show that at least one random parameter must appear in the decryption, meaning that either $x_n$ must appear or $S_{E \cap D}$ is non-empty (or both).
Assume this is not the case. Then we have
\AR{
\text{XOR}_D(a,b,\mathbf{x},m_s) = \text{XOR}'_D(a,b,m_s) 
}
and this must be equal to $ab$ by the correctness of the protocol.
But, due to the restrictions of the client we have
\AR{
\text{XOR}'_D(a,b,m_s)=\text{XOR}''_D(a,b) \oplus m_s =ab \ \textup{or} \\ \text{XOR}'_D(a,b,m_s)=\text{XOR}''_D(a,b) = ab
}
The latter is not possible as no function computable using only XOR can yield the output $ab$, so
\AR{
\text{XOR}'_D(a,b,m_s)=\text{XOR}''_D(a,b) \oplus m_s =ab \Leftrightarrow\\
m_s =ab \oplus \text{XOR}''_D(a,b).}

The function $\text{XOR}''_D(a,b)$ can only be one of six functions, which are such that either $a$ or $b$ appear in the decryption:
\AR{
\text{XOR}''_D(a,b) = a;\ \text{XOR}''_D(a,b) = 1\oplus a\\
\text{XOR}''_D(a,b) = b;\ \text{XOR}''_D(a,b) = 1\oplus b;  \\
\text{XOR}''_D(a,b) = a \oplus b;\ \text{XOR}''_D(a,b) = 1 \oplus a \oplus b.
}
But, for all of these functions we have that $ab \oplus \text{XOR}''_D(a,b)$ is correlated to $a,b$, hence not blind.
For example $ a \oplus b \oplus ab = a\vee b,$ so if the server obtains $m_s=0$ this means $a=b=0.$ 
Thus, for the protocol to be blind, at least one random parameter must appear in the decryption.

Let $j$ be the index of this random parameter. Then $x_j$ either appears or does not appear in the encryption.
First assume $x_j$ appears in the encryption, and let 
$\text{XOR}''_D(a,b) = a.$ Then by modifying $\text{XOR}_D$ in such a way that it no longer depends on $a$ (by substituting $\text{XOR}''_D(a,b)$ with $0$ in the definition of $\text{XOR}_D$) and by modifying the encryption function in such a way that all 
instances of $x_j$ are substituted with $x_j \oplus \text{XOR}''_D(a,b)$, we obtain a new protocol, in which the inputs $a,b$ no longer appear in the decryption function.
This protocol is correct, as the initial protocol was correct for all possible inputs and random variables, and all we have done is a substitution of variables.
Since, from the perspective of the server, both $x_j \oplus \text{XOR}''_D(a,b)$ and $x_j$ are equally distributed (uniformly at random), the protocol is blind as well.

Consider now the case where $x_j$ does not appear in the encryption (thus no random parameters appearing in the encryption appear in the decryption), and let $\text{XOR}''_D(a,b)$ be the function which appears in the evaluation of the decryption, and is not constant. 
Then, we need to modify the messages the client sends, and the measurement the server does.
Let $m_c$ be the message the client sends in the original protocol.
Then, in the modified protocol, the client will send the message $(m_c, \text{XOR}''_D(a,b) \oplus y)$, where $y$ is a new random bit.
The server will perform the same measurement as in the original protocol, as defined by $m_c$ but will output
$m_s^{new} = m_s^{original} \oplus \text{XOR}''_D(a,b) \oplus y$.
Note that this process can be viewed as a redefinition of the measurement the server does.
the client decrypts {almost the same} as in the original protocol, {altered} {by} substituting $\text{XOR}''_D(a,b)$ with $0$, and by XORing with $y$ .
So we have:
\AR{
\textup{The\ original\ decryption\ in\ original\ protocol}:\\
out = \text{XOR}''_D(a,b) \oplus m_s^{original} \oplus x_j \\
\textup{The\ new\ decryption\ in\ new\ protocol}:\\
0 \oplus m_s^{new}  \oplus x_j \oplus y = m_s^{original} \oplus \text{XOR}''_D(a,b) \oplus y \oplus x_j \oplus y = out.
}
Thus, the new protocol is also correct. To see that it is blind, note that the only piece of additional information given to the server, relative to the original protocol is the bit $\text{XOR}''_D(a,b) \oplus y.$ However, since $y$ is chosen uniformly at random, this reveals no extra information so the protocol is blind as well.

Thus for every QO2$_3$ blind correct protocol, there exists a blind correct QO2$_4$ protocol where the inputs of the client do not appear in the decryption function. \qed

To summarise, to this point we have shown that we only need to consider protocols in which the server's output is a single bit, at most one random parameter which appears in the decryption (but not in encryption) is {used}, and the decryption function does not take the inputs of the client as parameters. Additionally we have shown that we only need the random parameters which appear either in encryption or decryption. Next, we deal with the size of the client's messages, and the number of required random parameters appearing in the encryption.

Consider the encryption, and the generated quantum state in the protocol:
\AR{
m_c = \text{XOR}_E (a,b, \mathbf{x})\ - \ \textup{the\ m\ bit\ message\ from\ the client}\\
\rho^{\mathbf{x}}, \ for\ \mathbf{x} = (x_1, \ldots, x_n)\ -\ \textup{the\ quantum\ state\ parametrized\ by\ } n\ \textup{bits}.\\
}
and let $(m_c)_j$ denote the $j^{th}$ bit of the $m$ bit message $m_c$.

\noindent {\bf Claim 5.} No single isolated random variables are needed. 
\proof
Assume that, for some $j$ and $k$ we have, $(m_c)_j = x_k$.
Then, the protocol reveals $x_k$. But this means that if we fix $x_k=0$ (that is, by dropping that random parameter from the protocol) we yield again a blind correct protocol (with one less random parameter). We get the same if the negation of $x_k$ appears. 
By repeating this, we obtain a protocol for which no part of the message is equal to a single random parameter, or its negation. \qed

\noindent {\bf Claim 6.} No arbitrary XOR functions of random variables are needed. 
\proof
Next, assume that for some $j$ and $k,l$ we have, $(m_c)_j = x_k \oplus x_l$.
Then, we can introduce the variable $x_{k,l} = x_k \oplus x_l$, and substitute all instances of $x_l$ in the protocol with $x_{k,l} \oplus x_l$. This again yields a correct blind protocol, with the same number of random parameters as the original protocol. However, the modified protocol has the new variable $x_{k,l}$ appearing in $(m_c)_j$ isolated, so it (by the argument in the last paragraph) be dropped from the protocol.

We can perform analogous substitutions whenever arbitrary XOR functions of random parameters appear in isolation:
for a function $b \oplus x_{k_1} \oplus \cdots \oplus x_{k_p}$ we can define the substituting variable 
$x_{k_1, \ldots, k_p}^b =b \oplus x_{k_1} \oplus \cdots \oplus x_{k_p}$, and substitute all instances of
$x_{k_1}$ with $x_{k_1, \ldots, k_p}^b \oplus b \oplus x_{k_2} \oplus \cdots x_{k_p}$. Thus we retain exactly the same number of random parameters, but $x_{k_1, \ldots, k_p}^b$ now appears in isolation. So, this variable can be dropped.

Thus, for any QO2$_4$ protocol, there exists a protocol (blind and correct) where no functions of random parameters appear in isolation in $m_c$.

Thus, each entry of $m_c$ is of the form $\text{XOR}(a,b,x_1, \ldots x_n)$, where this function is not constant in $a$ or $b$ (or both). However, it is clear that this function cannot be constant in all the random parameters $\mathbf{x}$ as otherwise the protocol would not be blind. \qed

We can now complete the main proof of the impossibility of quantum offline protocol by showing how the redundancies could be removed.

\noindent {\bf Proof of Theorem \ref{t-qoffline}}

Assume $(m_c)_j = \text{XOR}(a,b) \oplus \bigoplus_{k \in S_j \subseteq \left[ N \right ]} x_k$ and 
$(m_c)_{k \not = j} = \text{XOR}(a,b) \oplus \bigoplus_{k \in S_k \subseteq \left[ N \right ]} x_k$ (that is the same function of $a,b$ appears twice in the message).
Then, the XOR of those two entries reveals the XOR of the random parameters with indices in the intersection $S_j \cap S_k$.
 Let \AR{
 \tilde{x} = \bigoplus_{k \in S_j \subseteq \left[ N \right ]} x_k \oplus  \bigoplus_{k \in S_k \subseteq \left[ N \right ]} x_k  = \bigoplus_{k \in S_k \cap S_j \subseteq \left[ N \right ]} x_l }
Then the original protocol is equally blind as
the protocol (we will call it MOD1 for modification 1) in which the message element $(m_c)_k$ is substituted with $\tilde{x}$ and the server, upon the receipt of the message redefines $(m_c)_k \df (m_c)_j \oplus x$.

For simplicity, assume that $S_k \cap S_j  = \{1,2, \ldots l \}$.
If we further modify MOD1 to MOD2 by substituting all instances of $x_1$ in this protocol with $\tilde{x} \oplus x_2 \ldots x_l$ we obtain a protocol in which $\tilde{x}$ is a randomly chosen variable, and note that it appears isolated in message element $(m_c)_k$.
Thus, it can by the arguments we presented earlier, be dropped from the protocol, by setting it to zero.
Note that analogous transformations of the protocol can be done if the XOR functions on two positions differ by a bit flip.

Hence, we only need to consider protocols where each function of $a,b$ in the message of the client appears only once, where functions which differ by a bit flip can be considered duplicates as well.
There are only three XOR computable non-constant functions of two binary parameters, up to a bit flip:
\AR{
\text{XOR}(a,b) = a, \; \text{XOR}(a,b)=b, \; \text{XOR}(a,b)=a\oplus b 
}
Thus, the message the client sends to the server, without the loss of generality, is of the form:
\AR{
m_c = (a \oplus \bigoplus_{k \in S_1 \subseteq \left[ n \right ]} x_k, b \oplus \bigoplus_{k \in S_2 \subseteq \left[ n \right ]} x_k,a\oplus b \oplus \bigoplus_{k \in S_3 \subseteq \left[ n\right ]} x_k)
}
Now, we can eliminate any single one of the three, and for our purposes of reduction to the small QO2 protocol, we will eliminate the last one.
Note that 
\AR{
(m_c)_3 = (m_c)_1 \oplus (m_c)_2 \oplus \bigoplus_{k \in S_1 \subseteq \left[ n \right ]} x_k \oplus \bigoplus_{k \in S_2 \subseteq \left[ N \right ]} x_k \oplus \bigoplus_{k \in S_3 \subseteq \left[ N \right ]} x_k,
}
and that the server can obtain
\AR{
\tilde{x} =\bigoplus_{k \in S_1 \subseteq \left[ n \right ]} x_k \oplus \bigoplus_{k \in S_2 \subseteq \left[ n \right ]} x_k \oplus \bigoplus_{k \in S_3 \subseteq \left[ n \right ]} x_k
}
by XORing the three bits of the client's message. Thus, similarly to the approach we used earlier, the protocol can be further modified in such a way that $\tilde{x}$ is given as the third bit of the message. Furthermore, by substitution, the third bit can be eliminated as well.
Thus we obtain the third modification of the protocol, in which the client's message is of the form
\AR{
m_c = (a \oplus \bigoplus_{k \in S_1 \subseteq \left[ n \right ]} x_k, b \oplus \bigoplus_{k \in S_2 \subseteq \left[ n \right ]} x_k)
}
with $S_1 \cup S_2 = \left [ n \right]$.
Note $S_1 \not = S_2$ as otherwise the protocol would not be blind.
Let $S_{DE}$ be the subset of indices of the random parameters which appear in the decryption and encryption.
Then all the random parameters in $S_1 \setminus (S_2 \cup S_{DE})$ can be substituted by only one random parameter $\tilde{x}_1$ which is the mod 2 sum of random parameters indexed in $S_1 \setminus (S_2 \cup S_{DE})$. Additionally, the quantum state the client sends to the server needs to be averaged over all states where the mod 2 sum of random parameters indexed in $S_1 \setminus (S_2 \cup S_{DE})$ is zero (for $\tilde{x}_1 = 0$) and for the case it is one (for $\tilde{x}_1 = 1$).
The same can be done for all the random parameters in $S_2 \setminus (S_1 \cup S_DE)$, generating the single random parameter $\tilde{y}_1$ appearing only in $(m_c)_2$. 

The indices in $S_{DE}$ must appear either in $S_1$ or in $S_2$. Let $p_1 \ldots p_q$ be the set which appears in both.
Then we can substitute these random parameters with one $\tilde{p} = p_1\oplus \cdots \oplus p_q$ by again modifying the state the client sends to the server, by averaging over those states for which $p=0$ or $p=1$.
Similarly can be done for those indices in $S_{DE}$ which appear only in $(m_c)_1$ (same for $(m_c)_2$ ) resulting in one random parameter $\tilde{x}_2$ ($\tilde{y}_2$).

Thus we obtain the protocol in which the client sends
\AR{
m_c = (a \oplus \tilde{x}_1  \oplus \tilde{x}_2  \oplus p, b \oplus \tilde{y}_1  \oplus \tilde{y}_2  \oplus p)
}
and the decryption is given with:
\AR{
out = m_s \oplus \tilde{x}_2 \oplus \tilde{y}_2 \oplus p \oplus r
}
where $r$ was the random parameter not appearing in the encryption, and the quantum state is parametrized with:
\AR{
\rho^{\tilde{x}_1, \tilde{x}_2 ,  \tilde{y}_1  , \tilde{y}_2,p,r}
}
We will refer to such protocols as QO2$_5$ protocols.

Note that 
\AR{
M^{\alpha, \beta} (\rho^{\tilde{x}_1, \tilde{x}_2 ,  \tilde{y}_1  , \tilde{y}_2,p,r})= (\alpha \oplus \tilde{x}_1 \oplus \tilde{x}_2 \oplus p)(\beta \oplus \tilde{y}_1 \oplus \tilde{y}_2 \oplus p) \oplus \tilde{x}_2 \oplus \tilde{y}_2 \oplus p \oplus r
}
and equivalently that
\AR{
M^{\alpha, \beta} (\rho^{\tilde{x}_1', \tilde{x}_2' ,  \tilde{y}_1' , \tilde{y}_2,p',r'})= (\alpha \oplus \tilde{x}_1' \oplus \tilde{x}_2' \oplus p')(\beta \oplus \tilde{y}_1' \oplus \tilde{y}_2' \oplus p') \oplus \tilde{x}_2' \oplus \tilde{y}_2' \oplus p' \oplus r'.
}
Therefore we obtain the following relation:
\AR{
M^{\alpha, \beta} (\rho^{\tilde{x}_1, \tilde{x}_2 ,  \tilde{y}_1  , \tilde{y}_2,p,r}) =M^{\alpha, \beta} (\rho^{\tilde{x}_1', \tilde{x}_2' ,  \tilde{y}_1'  , \tilde{y}_2',p',r'}) \ if\   \\
\tilde{x}_1 \oplus \tilde{x}_2 \oplus p=\tilde{x}_1' \oplus \tilde{x}_2' \oplus p', \ and  \\
\tilde{y}_1  \oplus \tilde{y}_2 \oplus p = \tilde{y}_1'  \oplus \tilde{y}_2' \oplus p' \ and \\
\tilde{x}_2 \oplus \tilde{y}_2 \oplus p \oplus r = \tilde{x}_2' \oplus \tilde{y}_2' \oplus p' \oplus r'. 
}

Since the state $\rho$ is parametrized by 6 independent parameters and we have three independent equations, this implies that there are 8 differing equivalency classes (as defined by the three equalities) over the set of all possible random parameters.
The equivalency classes can be represented by three bits $c_1, c_2, c_3$ as follows:
\AR{
(c_1,c_2,c_3) \equiv \left\{ (\tilde{x}_1, \tilde{x}_2 ,  \tilde{y}_1  , \tilde{y}_2,p,r) \vert \tilde{x}_1 \oplus \tilde{x}_2 \oplus 
p = c_1 \right.\\
 \left. \tilde{y}_1  \oplus \tilde{y}_2 \oplus p=c_2,  \tilde{x}_2 \oplus \tilde{y}_2 \oplus p \oplus r = c_3   \right\}
}
We can then define the states $\rho$, averaged per equivalency class:
\AR{
\rho^{c_1, c_2,c_3} = 1/8 \sum_{(\tilde{x}_1, \tilde{x}_2 ,  \tilde{y}_1  , \tilde{y}_2,p,r) \in (c_1, c_2,c_3)} \rho^{\tilde{x}_1, \tilde{x}_2 ,  \tilde{y}_1  , \tilde{y}_2,p,r}
}
Note that the first bit of the message the client sends to the server in QO2$_5$ is given with $(a\oplus x_1 \oplus x_2 \oplus p)$ which is equal to $c_1$. Similarly, the second bit $(b\oplus y_1 \oplus y_2 \oplus p)$ is equal to $c_2$.
The decryption is given with $out = m_s \oplus x_2 \oplus y_2 \oplus p \oplus r$ which is equal to $m_s \oplus c_3$.

This gives us a protocol in which:
 the client sends
\AR{
m_c = (a \oplus c_1, b \oplus c_2)
}
and the decryption is given with:
\AR{
out = m_s \oplus c_3
}
where $c_3$ was the random parameter not appearing in the encryption, and the quantum state is parametrized with:
\AR{
\rho^{c_1, c_2,c_3}
}
This protocol is correct by construction, and it is also blind as the classical messages the client sends are the same as in the QO2$_5$ protocol, and the quantum state is averaged over the degrees of freedom which do not appear in the abbreviated protocol - but then the averaging over the remaining free parameters yields the same state on the server's side as in the QO2$_5$ protocol. Thus it is blind as well.

But this is also a small QO2 protocol.
Thus, symbolically, we have shown:

\AR{
\exists \textup{QO2} \rightarrow \exists \textup{QO2}_1 \rightarrow \exists \textup{QO2}_2\rightarrow \exists \textup{QO2}_3\rightarrow \exists \textup{QO2}_4\rightarrow \exists \textup{QO2}_5 \rightarrow \exists \textup{ small \ QO2}
}
which implies the proof of the main theorem since we have already proven no small QO2 protocol exists. \qed

{We believe that} multiple rounds of classical or quantum communication cannot help either. This can be seen as the operation the client tries to perform, that is an AND, cannot be broken down into a sequence of operations which are themselves not universal for classical computation, when used in conjunction with XOR and NOT. {However, we leave the general proof of impossibility for future work.}

The above discussion also points at the impossibility of extending our protocol to entangled non-commuting servers to remove any quantum component from the client side. This scenario was originally proposed for the universal Blind Quantum Computing \cite{UBQC} where the client requests one server to measure his part of entangled state in the basis $1/ \sqrt 2(\ket{0} \pm e^{i\theta \ket 1})$, $\theta$ being a randomly chosen parameter known only to server 1 and the client. This step replace the requirement of the client device to prepare and send single qubit to the server 2. Now, the client could follow the step of the original protocol by adapting the required correction due to the random result of the measurement of the server 1. In the above construction revealing parameter $\theta$ to server 1 does not effect the blindness \cite{UBQC}. However as proved before, any blind quantum AND protocol requires quantum states that are dependent on the client's input. Therefore once could not delegate the preparation of such states to any untrusted servers. 

\section {Discussion}

The family of SecureNAND protocols presented in this paper highlights the role of a single quantum state in obtaining a security task  unattainable in a classical setting, mirroring the super-dense coding protocol \cite{BW92} for communication tasks. While the presented no-go results emphasize new conceptual aspects of quantum theory and could potentially be linked to the study of quantum games, a new exciting direction we envision to explore further is a hybrid quantum-classical scheme for delegated classical computing. Any advance{ment} to the problem of secure delegated computation would have an immediate significant consequence on how computational problems are solved in the real
world. One can envision virtually unlimited computational power to end users on the go, using just a simple terminal to access the computing cloud which would turn any smartphone into a Òquantumly--enhanced smartphoneÓ. Only then could we truly justify our proposed title of quantum-enhanced secure delegated classical computing!  While the crucial challenge in developing classical schemes for delegated computing is the design of encryption procedures that are independent from the complexity of the function of interest, in SecureNAND protocols the bottleneck is the required quantum communication between the client and the server. These two seemingly unrelated features are in fact deeply connected as our no-go results demonstrate and the investigation of their relationship will dictate the practical success of a possible hybrid quantum-classical delegated computing.

\bibliographystyle{unsrt}
\bibliography{bibliography}

\appendix
\section{Single qubit-based Protocols} 

\begin{figure*}[h!] 
   \centering
   \includegraphics[width=300pt]{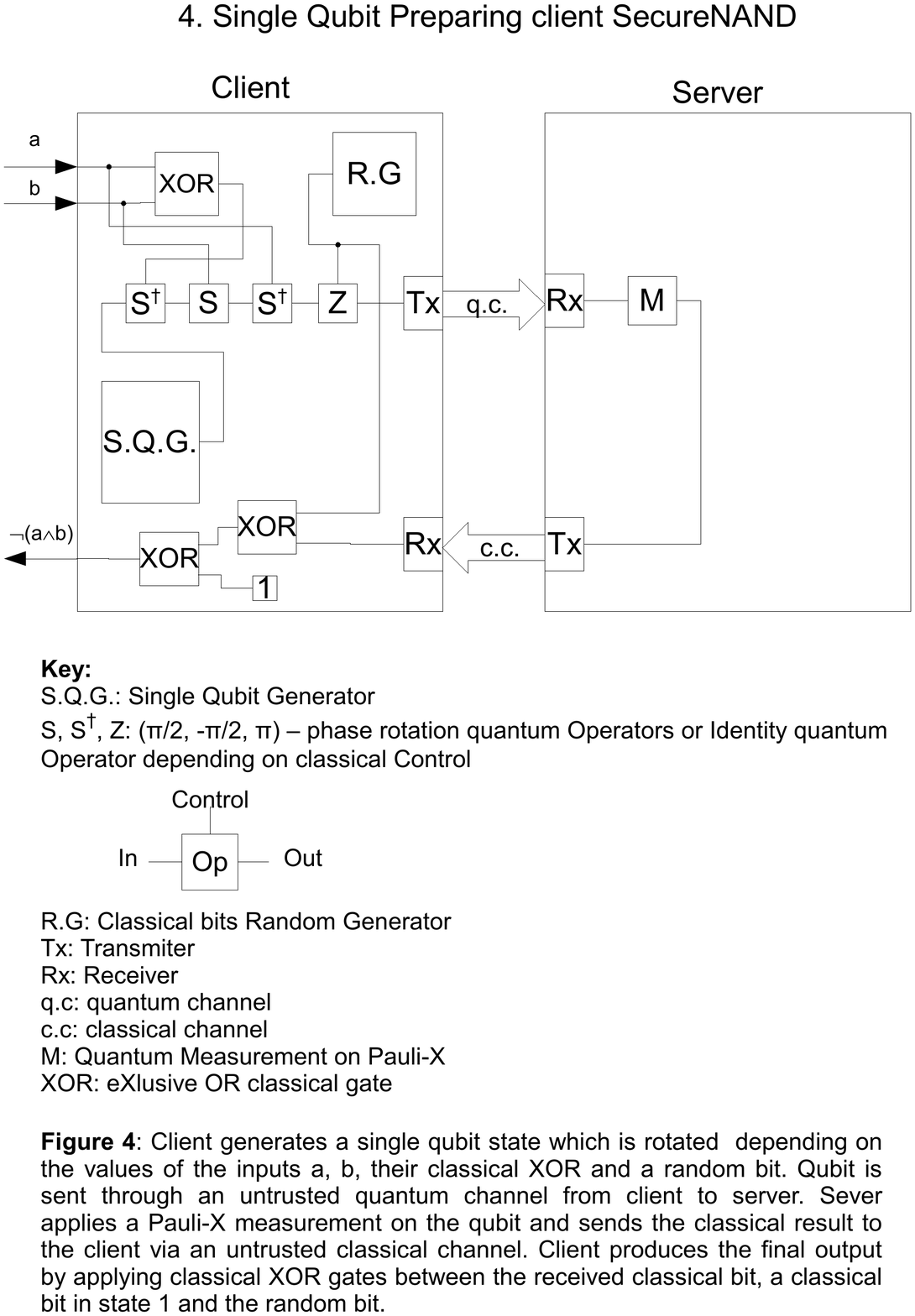} 
   \label{Fig-Implement4}
\end{figure*}

\begin{figure*}[h!] 
   \centering
   \includegraphics[width=300pt]{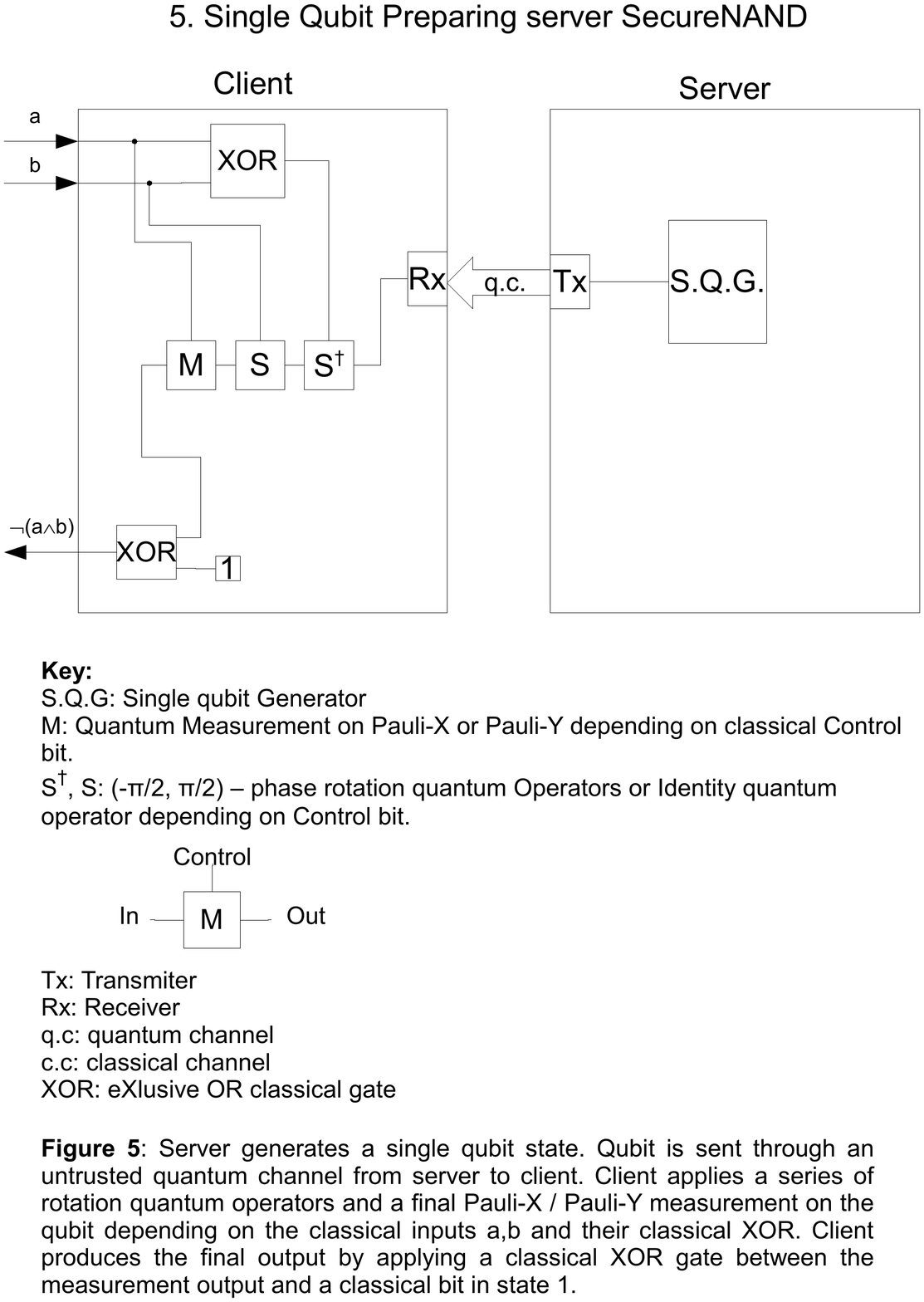} 
   \label{Fig-Implement5}
\end{figure*}

\begin{figure*}[h!] 
   \centering
   \includegraphics[width=300pt]{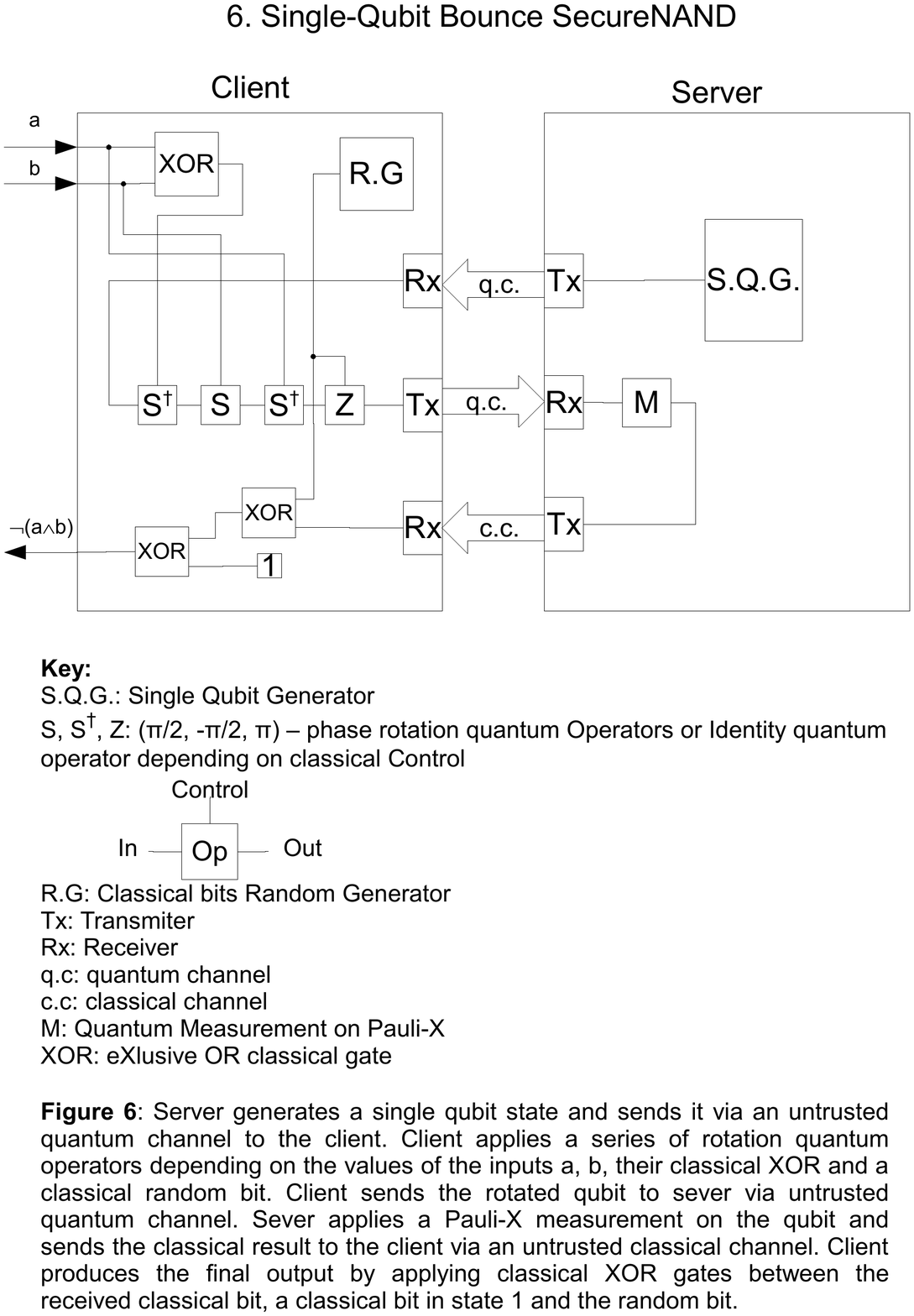} 
   \label{Fig-Implement6}
\end{figure*}
\end{document}